\documentclass[fleqn,usenatbib]{mnras}

\usepackage{newtxtext,newtxmath}
\usepackage[T1]{fontenc}
\usepackage{ae,aecompl}

\usepackage{amssymb}
\usepackage{amsmath}
\usepackage{graphicx}
\usepackage{ulem}
\usepackage{multirow}
\usepackage{diagbox}
\usepackage{natbib}

\def\gsim{\;\rlap{\lower 2.5pt\hbox{$\sim$}}\raise 1.5pt\hbox{$>$}\;}
\def\lsim{\;\rlap{\lower 2.5pt\hbox{$\sim$}}\raise 1.5pt\hbox{$<$}\;}
\def\dd{\mathrm{d}}
\def\Om{\Omega_\mathrm{m}}
\def\bfc{\mathbf{c}}
\def\bfe{\mathbf{e}}
\def\bfgamma{\boldsymbol{\gamma}}
\def\bftheta{\boldsymbol{\theta}}
\def\bfell{\boldsymbol{\ell}}

\title[$\sigma_8$ inferred from power spectrum and 2PCF]{The matter fluctuation amplitude inferred from the weak lensing power spectrum and correlation function in CFHTLenS data}

\author[T. Lu \& Z. Haiman]{
Tianhuan Lu $^{1}$\thanks{E-mail: tl2854@columbia.edu}
and Zolt\'an Haiman $^{1}$ \\
$^{1}$Department of Astronomy, Columbia University, New York, NY 10027, USA
}

\pubyear{2019}

\begin{document}
\label{firstpage}
\pagerange{\pageref{firstpage}--\pageref{lastpage}}
\maketitle

\begin{abstract}
Based on the cosmic shear data from the Canada-France-Hawaii Telescope
Lensing Survey (CFHTLenS), \citet{kilbinger2013} obtained a constraint
on the amplitude of matter fluctuations of
$\sigma_8(\Om/0.27)^{0.6}=0.79\pm0.03$ from the two-point correlation
function (2PCF). This is $\approx3\sigma$ lower than the value
$0.89\pm0.01$ derived from {\it Planck} data on cosmic microwave
background (CMB) anisotropies.  On the other hand, based on the same
CFHTLenS data, but using the power spectrum, and performing a
different analysis, \citet{liu2015} obtained the higher value of
$\sigma_8(\Om/0.27)^{0.64}=0.87^{+0.05}_{-0.06}$.  We here investigate
the origin of this difference, by performing a fair side-by-side
comparison of the 2PCF and power spectrum analyses on CFHTLenS
data. We find that these two statistics indeed deliver different
results, even when applied to the same data in an otherwise identical
procedure. We identify excess power in the data on small scales
($\ell>5,000$) driving the larger values inferred from the power
spectrum. We speculate on the possible origin of this excess
small-scale power. { More  generally, our results highlight the
utility of  analysing the 2PCF and the power spectrum in tandem,
to discover (and to help control) systematic errors.}
\end{abstract}


\section{Introduction}
\label{sec:intro}

Weak gravitational lensing (WL) is a promising probe of the mass
distribution in the Universe, via observations of correlated
distortions in the shapes of background galaxies~\citep[e.g.][]{bs2001,
 refregier2003, hj2008, kilbinger2015}.
Weak lensing has also matured as a method to measure the
parameters of the background $\Lambda$ cold dark matter ($\Lambda$CDM)
cosmological model.
Several recent WL surveys have yielded competitive constraints on cosmological
parameters, including the Canada-France-Hawaii Telescope Lensing Survey (CFHTLenS;
\citealt{heymans2012, kilbinger2013}), the Kilo-Degree Survey (KiDS;
\citealt{kuijken2015, hildebrandt2016, kohlinger2017}), the Dark Energy
Survey (DES; \citealt{abbott2016, troxel2018}), and the Subaru Hyper
Suprime-Cam (HSC; \citealt{aihara2017a, aihara2017b, hikage2019}) survey.

As the size and depth of weak lensing surveys have increased, a
moderate tension has emerged on parameter constraints, especially on
the matter density fluctuation amplitude $\sigma_8$, between the
results from weak lensing and from the cosmic microwave background
(CMB). Specifically, Planck 2018 \citep{planck2018}, based on the CMB,
inferred the joint constraint on $\sigma_8$ and the matter density
$\Om$, $\Sigma_8\equiv\sigma_8(\Om/0.3)^\alpha=0.83\pm0.01$ for
$\alpha=0.5$ (or equivalently $\sigma_8(\Om/0.27)^{0.6}=0.89\pm0.01$).
\citet{kilbinger2013}, based on WL data in CFHTLenS, found
$\sigma_8(\Om/0.27)^{0.6}=0.79\pm0.03$, which is lower than the Planck
value by $\sim3\sigma$ (note: all error-bars correspond to 68\%
confidence). Similar discrepancies have been identified between other
WL surveys \citep[e.g.][]{hildebrandt2016, kohlinger2017, hikage2019},
and Planck's earlier results \citep{planck2015}.

An exception to this trend is the analysis of the CFHTLenS lensing
data by \citet[][hereafter L15]{liu2015}, who have found a higher
fluctuation amplitude, close to the Planck value.  L15 performed a set
of $N$-body simulations on a grid of 91 cosmological models with a
range of $\Om$, $\sigma_8$, and $w$ values (where $w$ is the equation
of state parameter for dark energy), and replicated the sky positions,
redshifts, and shape noise of $\sim4\,\textrm{million}$ galaxies from
the public CFHTLenS data. They fit the simulations to the observed
data using two different statistics: the convergence power spectrum,
and the abundance of lensing peaks (defined as local maxima on the
convergence maps).  They inferred
$\sigma_8(\Om/0.27)^{0.64}=0.87^{+0.05}_{-0.06}$ and
$0.84^{+0.04}_{-0.03}$, respectively, from these statistics.

These results are especially interesting, because
\citet{kilbinger2013} and L15 are based on the same WL survey and use
a similar pipeline for parameter estimation, but their constraints
differ at the $~2\sigma$ significance. We also note that
\citet{kilbinger2013} used a different statistic -- the two-point
correlation function (2PCF) of the WL shear -- which could cause a
discrepancy.  These two statistics have yielded different values in
the KiDS data, with \citet{hildebrandt2016} finding
$\sigma_8(\Om/0.3)^{0.5}=0.745\pm0.039$ using the 2PCF, and
\citet{kohlinger2017} finding the {\it lower} value
$\sigma_8(\Om/0.3)^{0.5}=0.651\pm0.058$ using the power spectrum.

In this paper, we focus on the difference between power spectrum and
2PCF, and explore the reasons for the discrepancy on $\Sigma_8$
between L15 and other weak lensing works, primarily
\citet{kilbinger2013}.
In~\S~\ref{sec:data} we summarise salient features of the CFHTLenS
data, and how we processed them.
In~\S~\ref{sec:methods}, we describe our methodology, including the
suite of ray-tracing simulations, and the likelihood analysis using
the two different statistics.
In~\S~\ref{sec:results}, we present our results, which reproduce the
previous discrepancy, and then test possible reasons for this
discrepancy.
We then discuss our results, and several other possible data-related
and physical explanations in \S~\ref{sec:discussion}.
Finally, we summarise our conclusions in \S~\ref{sec:conclusion}.

\section{Lensing Data}
\label{sec:data}

\subsection{CFHTLenS shear catalogue}
\label{sec:catalogue}

The CFHTLenS data covers four separate patches on the sky,
W1$\rightarrow$W4, consisting of 179 pointings in total, and covering
a total survey area of $154\,\mathrm{deg}^2$. The CFHTLenS Catalogue,
which includes specifications of all objects identified in their data,
is publicly available \citep{erben2013}. In this work, we use the
galaxy shear data from the Catalogue, which was constructed with a
pipeline consisting of several steps: (1) extract a galaxy catalogue
by SExtractor \citep{ba1996}, (2) estimate photometric redshifts with
a Bayesian photo-$z$ code \citep{benitez2000}, and (3) measure galaxy
ellipticity values $\{e_1, e_2\}$, including multiplicative ($m$) and
additive ($c_2$) bias corrections calibrated with {\it lens}fit
\citep{miller2013}.

Following L15, we filter the shear catalogue and keep only galaxies in
the redshift range $0.2\leq z\leq 1.3$ and weight $w>0$ (the latter is
assigned to each galaxy by {\it lens}fit, based on the significance of
its ellipticity measurement), leaving $\sim4.2$ million galaxies. We
then project and rearrange the survey area into 13 square subfields
with a size of $12\,\deg^2$ each, for further map generation.

In order to mitigate the impact of systematic errors,
\citet{heymans2012} computed a systematics test parameter $U$ for each
pointing, and performed a selection based on this parameter. As a result of
this procedure, 129 unique pointings are marked
as having passed the systematics selection, effectively keeping
$\sim75\%$ of the data ($\sim3.2$ million galaxies). We include this
selection for our results in \S~\ref{sec:results}, and discuss its
impact in \S~\ref{sec:selection}.

\subsection{Convergence maps}

In this section, we describe our steps to prepare convergence maps, on
which the power spectrum calculations are performed. The lensing
convergence ($\kappa$) represents a distance-weighted projection of
the over-density along the line of sight. It cannot be measured from
galaxy shapes directly, but it can be inferred from the lensing shear
$\bfgamma$ \citep{ks1993}, which, in turn, can be estimated from a
smoothed galaxy ellipticity field, using the weak lensing
approximation.

\subsubsection{Identifying low galaxy number density regions}

Due to bright stars and bad pixels, some part of the survey area is
unusable for measuring galaxy shapes. These regions can be found in
the CFHTLenS Catalogue by checking the local galaxy number
density. Following L15, we identify these low number density regions
by a criterion of $n_\mathrm{gal}<5\,\mathrm{arcmin}^{-2}$ over a
smoothing scale of $1\,\mathrm{arcmin}$. \citet{liu2014} found that
the magnification bias introduced by such a selection process is
negligible for CFHTLenS. Among the 13 square subfields, this criterion
gives a good separation between normal and low-density regions, with
average number densities of
$n_\mathrm{normal}=9.1\,\mathrm{arcmin}^{-2}$ and
$n_\mathrm{low}=0.7\,\mathrm{arcmin}^{-2}$ respectively.

\subsubsection{Creating shear maps}
\label{sec:shear-maps}

In creating shear maps, normal and low-density regions contribute
differently, so they are represented separately.  In the
normal-density regions, individual galaxies contribute in a discrete
manner. Because galaxies have four properties, we define four fields
as follows to characterize them:
\begin{equation}
\label{eqn:convert-def}
\mathcal{X}(\bftheta)=\sum_g^N{\delta(\bftheta-\bftheta_g)\,\mathcal{X}_g},
\end{equation}
\begin{equation}
\mathcal{X}\textrm{ is one of }\{\bfe, \bfc, m, w\}
\end{equation}
where $\delta(\cdot)$ is the Dirac delta function, $g=1\rightarrow N$
the indices of individual galaxies, $\bfe_g$ the two-component
ellipticity, $m_g$ and $\bfc_g=\{0,c_2\}$ the multiplicative and
additive biases, $w_g$ the weight.  { Note that
  $\mathcal{X}_g$ in eq.~\eqref{eqn:convert-def} represents one of the
  dimensionless quantities of an individual galaxy, whereas
  $\mathcal{X}(\bftheta)$ represents a field with a dimension of
  $(\textrm{solid angle})^{-1}$, inheriting the dimension of the Dirac
  delta function.  }

In the low-density regions, we ignore the contribution from the 
galaxies and use zero shear with average weight density in a 
continuous manner as follows,
\begin{equation}
w_\mathrm{low}(\bftheta)=\left\{
\begin{array}{ll}
  \langle w\rangle\cdot n_\mathrm{normal}, & \textrm{in normal-density regions}\\
  0, & \textrm{in low-density regions}
\end{array}
\right. ,
\end{equation}
where $\langle w\rangle$ is the average weight of the galaxies and 
$n_\mathrm{normal}=9.1\,\mathrm{arcmin}^{-2}$ is the overall number 
density of galaxies in normal-density regions. This
method is based on the assumption that the low-density regions have
similar densities of galaxies as in normal regions but they are not
observed, and their expected ellipticity is zero.

We calculate shear maps by applying a Gaussian smoothing to the galaxy
ellipticity fields with respect to the weight fields:
\begin{equation}
\bfgamma(\bftheta)=\int\frac{W_\mathrm{G}(\bftheta-\bftheta')w(\bftheta')[\bfe(\bftheta')-\bfc(\bftheta')]\,\dd\bftheta'}{W_\mathrm{G}(\bftheta-\bftheta')\{w(\bftheta')[1+m(\bftheta')]+w_\mathrm{low}(\bftheta')\} },
\label{eqn:smoothing}
\end{equation}
where we use a Gaussian kernel of scale $\theta_\mathrm{G}=0.5\,\mathrm{arcmin}$
\begin{equation}
W_\mathrm{G}(\bftheta)=\frac{1}{2\pi\theta_\mathrm{G}^2}\exp\left(-\frac{|\bftheta|^2}{2\theta_\mathrm{G}^2}\right).
\end{equation}

\begin{figure}
\center
\includegraphics[width=7.5cm]{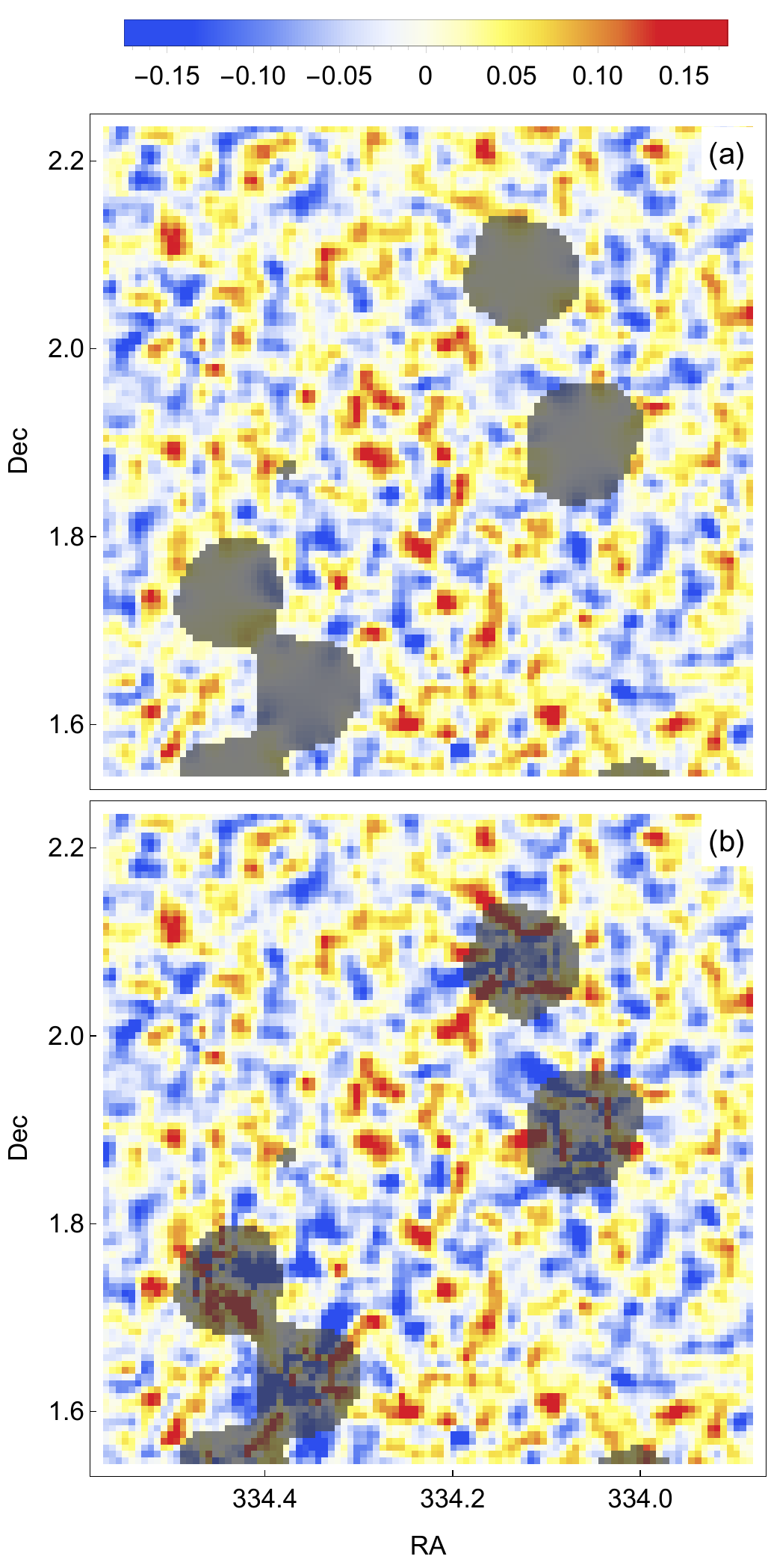}
\caption{Example portions of convergence maps generated (a) in this
  work and (b) by the method in L15.  Regions of low galaxy number
  density are marked by darker colours. Ignoring the weights of
  unobserved galaxies from low-density regions during the smoothing
  (see eq.\eqref{eqn:smoothing}) causes exceptionally large
  fluctuations near the boundaries on the scale of
  $1\sim10\,\mathrm{pixels}$. }
\label{fig:wlow-effect}
\end{figure}

In L15, the smoothing process was the same except they did not
consider the contribution from low density regions (effectively used
$w_\mathrm{low}(\bftheta)=0$). We noticed that doing so lets the shear
estimation in low-density regions be dominated by the shape noise from
very few galaxies. As shown in Figure~\ref{fig:wlow-effect}, setting
$w_\mathrm{low}(\bftheta)=0$ leads to high-frequency components near the
boundary, because of the abrupt changes in the number
density. This difference between L15 and our work significantly
affects the power spectrum on small scales (we discuss the
statistical impact of this effect in more detail in
\S~\ref{sec:power-spectrum} below).

\subsubsection{Creating convergence maps}
\label{sec:maps}

The lensing field can be expressed in terms of the lensing convergence
$\kappa$ and the complex-valued lensing shear
$\bfgamma=\gamma_1+i\gamma_2$. These two fields are related to the
lensing potential field $\psi$ by
\begin{eqnarray}
\kappa(\bftheta)&=&\frac{1}{2}\nabla^2\psi(\bftheta),\\
\gamma_1(\bftheta)&=&\frac{1}{2}(\psi_{,11}-\psi_{,22}),\\
\gamma_2(\bftheta)&=&\psi_{,12},
\end{eqnarray}
where
\begin{equation}
\psi_{,ij}\equiv\frac{\dd^2\psi}{\dd\theta_i\dd\theta_j}.
\end{equation}
Following \citet{ks1993}, the convergence field is related to the
shear field in Fourier space by
\begin{equation}
\hat{\kappa}(\bfell)=\frac{(\ell_1^2-\ell_2^2)\hat{\gamma_1}(\bfell)+2\ell_1\ell_2\hat{\gamma_2}(\bfell)}{\ell_1^2+\ell_2^2},
\end{equation}
where $\hat{\kappa}(\bfell)$ and $\hat{\gamma}_{1,2}(\bfell)$ denote
the Fourier transform of $\kappa(\bftheta)$ and
$\gamma_{1,2}(\bftheta)$ respectively. The Fourier transform and its
inverse are performed on a $512\times512$ grid for each
$12\,\mathrm{deg}^2$ square subfield. The pixel size
$\sim0.4\,\mathrm{arcmin}$ corresponds to a multipole of
$\ell\sim53,000$. A small portion of the convergence maps created in
this work is shown in panel (a) in Figure~\ref{fig:wlow-effect} for
illustration.

\subsection{Mock catalogues and maps}
\label{sec:mock}

In this work, we use the same mock galaxy shear catalogues as L15. We
here briefly introduce the steps of generating these mock catalogues, and
refer the reader to L15 for more details.

L15 picked 91 sampling points in a three-dimensional cosmological
parameter space: $0<\Om<1$, $-3<w<0$, and $0.1<\sigma_8<1.5$, where
the Latin hypercube sampling \citep{mckay1979} was used to make the
points statistically evenly distributed. They ran one $N$-body
simulation at each point using a modified version of the Gadget-2 code
\citep{springel2005}. $N$-body simulations were performed with $512^3$
dark matter particles in boxes of size $(240h^{-1}\mathrm{Mpc})^3$
starting from redshift $z=100$, with the additional cosmological
parameters set to constant values of $h=0.72$ (Hubble constant),
$\Omega_\mathrm{b}h^2=0.0227$ (baryon density), and
$n_\mathrm{s}=0.96$ (spectral index of scalar perturbations).

L15 used ray-tracing to create mock galaxy shears. At the sky position
of each galaxy in the CFHTLenS Catalogue, they followed a light ray
from redshift zero to the estimated galaxy redshift, and calculated
its lensing shear based on the gravitational potential. The $N$-body
simulation box for each cosmological model was randomly shifted and
rotated to create 1,000 pseudo-independent random realizations.

Note that this procedure places galaxies at their
observed locations, uncorrelated with the simulated field.  In
reality, the galaxies will reside in dark matter halos which are
correlated with the nearby density field, introducing possible
biases. However, we expect these biases to be negligible at the
sensitivity of CFHT, given that the density field at the source
plane is only weakly correlated with structures at the much
lower-redshift peak of the lensing kernel (see~\citet{liu2014} for
related discussion).

The lensing shear $\bfgamma$ in the mock catalogue so far contains no
noise. Following L15, we rotate the ellipticity of each individual
galaxy in the real CFHTLenS Catalogue by a random angle, and take this
to be the intrinsic ellipticity of that galaxy.  This is justified by
the fact that the r.m.s. cosmic shear signal contributes only of order
1\% of the observed total ellipticity.  We finally obtain the
ellipticity in the mock catalogue, using the weak lensing approximation,
as
\begin{equation}
\bfgamma_\mathrm{mock}=\bfe_\mathrm{mock}=
(1+m)\left(\bfgamma+\mathcal{R}(\varphi)\bfe\right),
\end{equation}
where $\bfe$ and $m$ are the ellipticity and multiplicative correction
of the real galaxy, $\mathcal{R}$ a 2-dimensional rotation matrix, and
$\varphi$ a random number drawn from a uniform distribution 
$\mathcal{U}(-\pi,\pi)$.

After generating the mock shear catalogues, we apply the same method
as described in \S~\ref{sec:maps} for the real data, to generate mock
convergence maps.  These shear catalogues and convergence maps are then
used to compute correlation functions and power spectra, as we describe
in the next section.

\section{Methods}
\label{sec:methods}

\subsection{Statistics}

We compare two statistics: the 2PCF of the shear, and the power
spectrum of the convergence field. We compute both statistic in each
square subfield individually, and then take the average over all
subfields for further estimation. The 2PCFs and the power spectra are
both computed over a finite range of angular scales. We chose the
minimum scale to be $1\,\mathrm{arcmin}$ ($\ell=21,600$) based on a map
pixel size of $\sim0.4\,\mathrm{arcmin}$. Considering that the linear
angular size of each square field is $\sim200\,\mathrm{arcmin}$, we
chose the maximum scale to be $60\,\mathrm{arcmin}$ ($\ell=360$),
which is consistent with L15.

In addition to the average value of both statistic, we make use of
1000 realizations of each cosmological model to calculate their
variance. Hereafter, we use $\xi_\pm(\vartheta)$ and $P(\ell)$ to
represent 2PCFs and power spectra, and use $\sigma_{\xi_\pm}$ and
$\sigma_P$ to represent the uncertainty of the mean values in terms of
their standard deviation.

\subsubsection{Two-point correlation function}

Following \citet{schneider2002}, the two components $\xi_+(\theta)$
and $\xi_-(\theta)$ of the 2PCF of the lensing shear field are
estimated by
\begin{equation}
\xi_\pm(\theta)=\frac{\sum_{ij}w_iw_j[e_\mathrm{t}(\bftheta_i)e_\mathrm{t}(\bftheta_j)\pm e_\times(\bftheta_i)e_\times(\bftheta_j)]}{\sum_{ij}w_iw_j}
\end{equation}
where $\theta=|\bftheta_i-\bftheta_j|$ is the angular scale, and
$e_\mathrm{t}, e_\times$ the tangential and cross components of the
galaxy ellipticity. The summation over all pairs of galaxies are
performed within $\theta$-bins evenly spaced on a logarithmic scale,
with the low-$\theta$ boundary of the $i$-th bin given by
$\theta_i=1.41^i\,\mathrm{arcmin}$, where $i=0,1,2,\cdots,11$.

\begin{figure}
\center
\includegraphics[width=7.5cm]{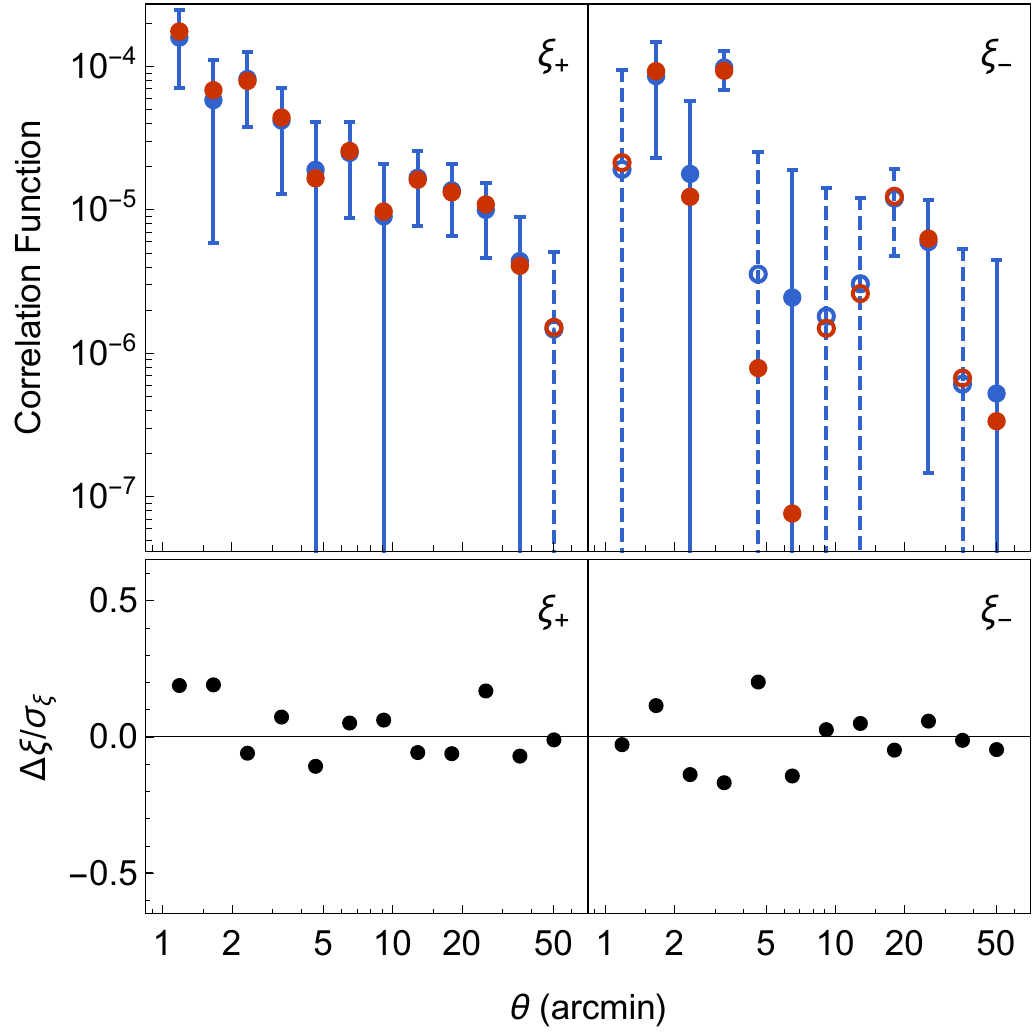}
\caption{The upper panel shows 2PCFs calculated by the brute-force
  method (red) and our faster grid-based method (blue).
  The error bars show the 68\% confidence interval of the
  correlation between ellipticities in each bin. Negative values
  are shown as empty circles/dashed error bars. The lower panel shows
  the residuals in units of the standard deviation. The calculations
  are performed on the ellipticities of the galaxies in the first
  subfield (RA $35.3^\circ\sim38.8^\circ$, Dec
  $-11.3^\circ\sim-7.8^\circ$).}
\label{fig:2pcf-calc}
\end{figure}

There are between $200-400$ thousand galaxies in each subfield, making
a summation over every pair of galaxies computationally
challenging. We wrote our own implementation of a code to calculate
the correlation function using a grid-based method, where we gradually
decrease the size of the cells to maintain accuracy at smaller and
smaller angular scale. The idea is similar to {\it Athena}
(see~\S~5~in \citealt{schneider2002}). The primary limitation of using
a grid-based method is that the angular resolution is lower. Our
method is implemented so that the minimum resolution is
$\Delta\theta/\theta=0.05$ at $\theta=1\,\mathrm{arcmin}$ and the
maximum resolution is $\Delta\theta/\theta=0.01$ for
$\theta\gsim5\,\mathrm{arcmin}$. As a reference, the resolution used
by \citet{kilbinger2013} was $\Delta\theta/\theta=0.03$ for all
scales.

Figure~\ref{fig:2pcf-calc} shows the result of our method compared to
the brute-force method.  While there are residuals, they are relatively small
$\Delta\xi/\sigma_\xi\lsim0.3$. Furthermore, we perform the
calculations on the real and the mock shear catalogues in the same way,
with only the numerical values of the ellipticities being different. 
Therefore, we do not expect a bias in our parameter
estimation, in spite of the existence of the small residuals.

\subsubsection{Power spectrum}
\label{sec:power-spectrum}

We perform an FFT of each convergence field on a
$512\times512$ grid and compute the power spectra in 38 logarithmic
multipole bins over the range $360<\ell<21,600$, where the
low-$\ell$ boundary of the $i$-th multipole bin is
$\ell_i=60^{i/38}\times360$, where $i=0,1,2,\cdots,37$.

\begin{figure}
\center
\includegraphics[width=7.5cm]{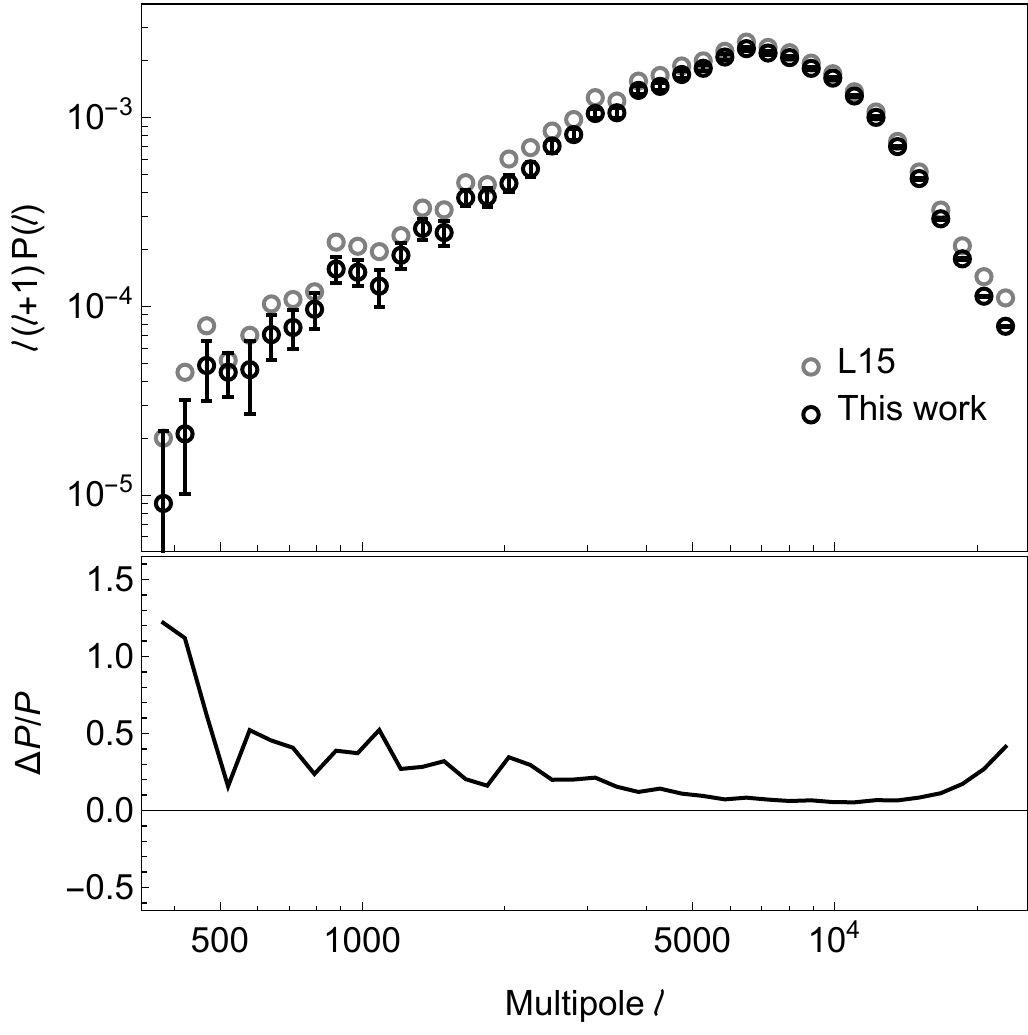}
\caption{The upper panel shows the power spectrum (with error bars
  corresponding to a standard deviation in each bin) calculated in
  this work, and that by L15's method, on the first subfield. The
  lower panel shows the relative differences between the two.}
\label{fig:spec-compare}
\end{figure}

In Figure~\ref{fig:spec-compare}, we compare the power spectrum we
derive from a subfield with that obtained by L15's method. L15 pointed
out two artefacts in their power spectra: (1) the finite pixel
size causes extra power at $\ell\gsim20,000$, and (2) hard cut-offs at
the edges of the masks, i.e. low number density regions, cause extra
power at $\ell\gsim7,000$. Compared to L15, we obtain a lower power
over all scales because we remove the second artefact by considering 
the contribution of low-density regions. 

\subsection{Parameter estimation}

We estimate the posterior probability distribution in the parameter
space $\mathbf{p}\in\{\Om,\sigma_8\}$ in a standard way, using Bayes'
theorem. Assuming a model $M$ and the observed data $\mathbf{d}$, the
posterior distribution is
\begin{equation}
p(\mathbf{p}|\mathbf{d},M)=\frac{p(\mathbf{d}|\mathbf{p},M)\,p(\mathbf{p}|M)}{p(\mathbf{d}|M)}. 
\end{equation}
We adopt uniform distributions on $0.1$$\leq$$\Om$$\leq$$0.8$ and
0.1$\leq$$\sigma_8$$\leq$$1.4$ for the prior $p(\mathbf{p}|M)$, and
the likelihood function is given by
\begin{equation}
\log p(\mathbf{d}|\mathbf{p},M)\propto-\left[\mathbf{d}-\mathbf{y}(\mathbf{p})\right]^\mathrm{T}\mathbf{C}^{-1}(\mathbf{p})\left[\mathbf{d}-\mathbf{y}(\mathbf{p})\right],
\end{equation}
where $\mathbf{y}(\mathbf{p})$ are the values of observables from the
$N$-body simulation and $\mathbf{C}$ the covariance matrix. Here, we
assume the statistics follow multivariate Gaussian distributions. In
general, $\mathbf{C}=\mathbf{C}(\mathbf{p})$ is a function of
cosmological model, but a constant covariance is often used in the
literature since it gives reasonable estimates. We use cosmology-dependent
covariances to obtain our primary results, and then compare these with
results using constant covariances in \S~\ref{sec:covariance}.

Finally, we need to interpolate from the 91 discrete cosmological
models to obtain the continuous function $\mathbf{y}(\mathbf{p})$. For
each of the 91 simulated models, we measure 2PCFs and power spectra
for all realizations, and find their averages over all realizations,
together with their covariances. Then, we interpolate between these
models to obtain the mean statistics and the covariances in the 3D
parameter space using polynomial fitting. Note that the mean
statistics and covariances are highly correlated among cosmological
models, so we can reduce their dimensions using principal component
analysis (PCA; \citealt{wold1987}) to avoid over-fitting.

\begin{figure}
\center
\includegraphics[width=7.5cm]{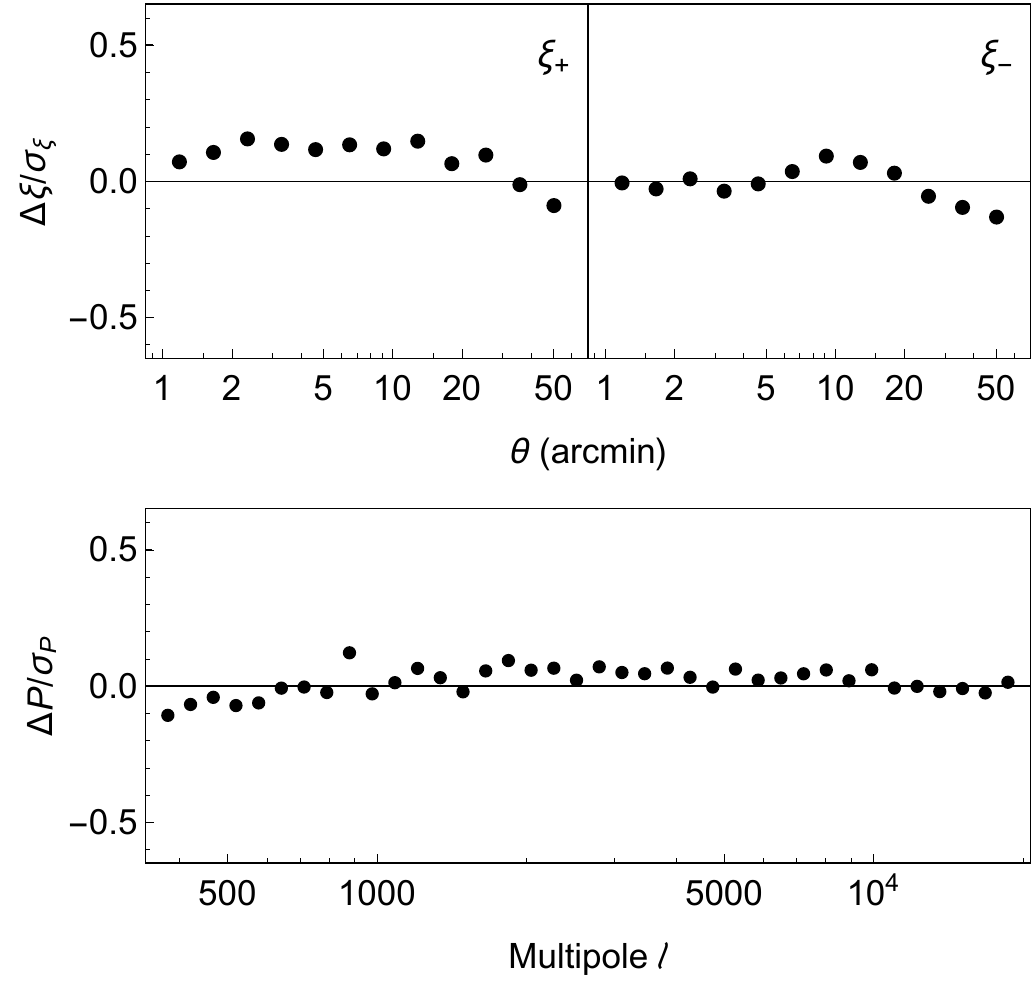}
\caption{The residuals between the interpolated 2PCFs and the power
  spectrum, relative to their true (directly measured) values. The
  residuals are divided by the corresponding standard deviation in
  each bin. The statistics are averaged over all 13 subfields.}
\label{fig:interpolation}
\end{figure}

After some experimentation and testing, we found the best performance
by setting the number of principal components $N_\mathrm{PC}=5$ and
fitting using polynomials up to 5th order. To illustrate the accuracy,
we pick a fiducial model ($\Om=0.305$, $\sigma_8=0.765$, $w=-0.879$)
closest to the Planck 2018 values, remove it from the polynomial
fitting dataset, and compare the predicted values with the true
values. As shown in Figure~\ref{fig:interpolation}, the residuals are
below $0.2\sigma$, i.e. much smaller than the
uncertainty for both statistics, giving us confidence that the
interpolation errors will not significantly affect parameter
estimation. Note that we interpolate along the direction of $w$, but
our analysis in this paper is restricted to the hyperplane $w=-1$,
corresponding to $\Lambda$CDM.

\section{Results}
\label{sec:results}

\subsection{Inferences from the 2PCF vs. the power spectrum}
\label{sec:verses}

We first reproduce the discrepancy between \citet{kilbinger2013} and L15 introduced only by the difference in methods, where the shear catalogues, $N$-body simulation, and data processing are exactly the same. 

\begin{figure}
\center
\includegraphics[width=7.5cm]{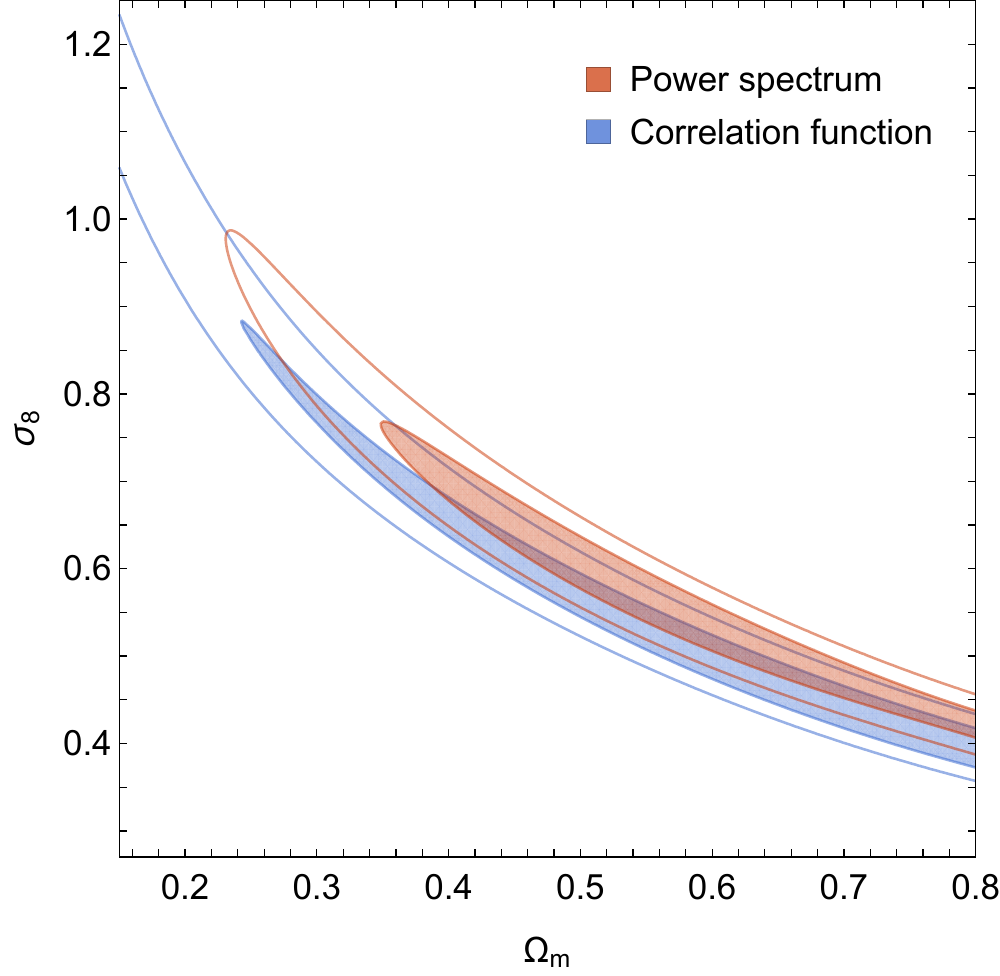}
\caption{68\% and 95\% joint confidence contours on $\sigma_8$ and
  $\Om$, using the power spectrum and the 2PCF, as labelled. The power
  spectrum yields an $\sim 8\%$ larger normalization, comparable to
  the $\sim10\%$ difference between L15 and \citet{kilbinger2013}.}
\label{fig:2pcf-vs-ps}
\end{figure}

Figure~\ref{fig:2pcf-vs-ps} shows confidence contours enclosing 68\%
and 95\% of the likelihood from the power spectrum and from the
2PCF. The parameter combination $\sigma_8(\Om/0.27)^\alpha$
(marginalised over the orthogonal direction in $[\sigma_8,\Om]$
parameter space) is best constrained with $\alpha=0.64$ for the 2PCF
and $\alpha=0.66$ for the power spectrum. We use a fixed $\alpha=0.65$
for simplicity, and we use $\Sigma_8\equiv\sigma_8(\Om/0.27)^{0.65}$
in order to do direct comparisons.

The best fits we find are $\Sigma_8=0.833\pm0.035$ using the 2PCF, and
$\Sigma_8=0.896\pm0.038$ using the power spectrum. Compared to the
$\sim10\%$ discrepancy on $\Sigma_8$ between \citet{kilbinger2013} and
L15, we find a comparable $\sim8\%$ difference, which is attributable
entirely to the use of a different statistic.

\subsection{Systematic biases}
\label{sec:biases}

In this section, we test whether there may be a systematic bias in
parameter estimates derived through either statistic.  To do this, we
constrain the cosmological parameters of a fiducial model from the
mock catalogues -- that is, we fit our own simulations, rather than
the observational data.  In each of the 1000 realizations of the
fiducial model ($\Om=0.305$, $\sigma_8=0.765$, $w=-0.879$) we create a
mock shear catalogue, and treat it as the real catalogue to derive
probability distributions in the $\Om$-$\sigma_8$ space, and to
compute the corresponding constraint on $\Sigma_8$.

\begin{figure}
\center
\includegraphics[width=7.5cm]{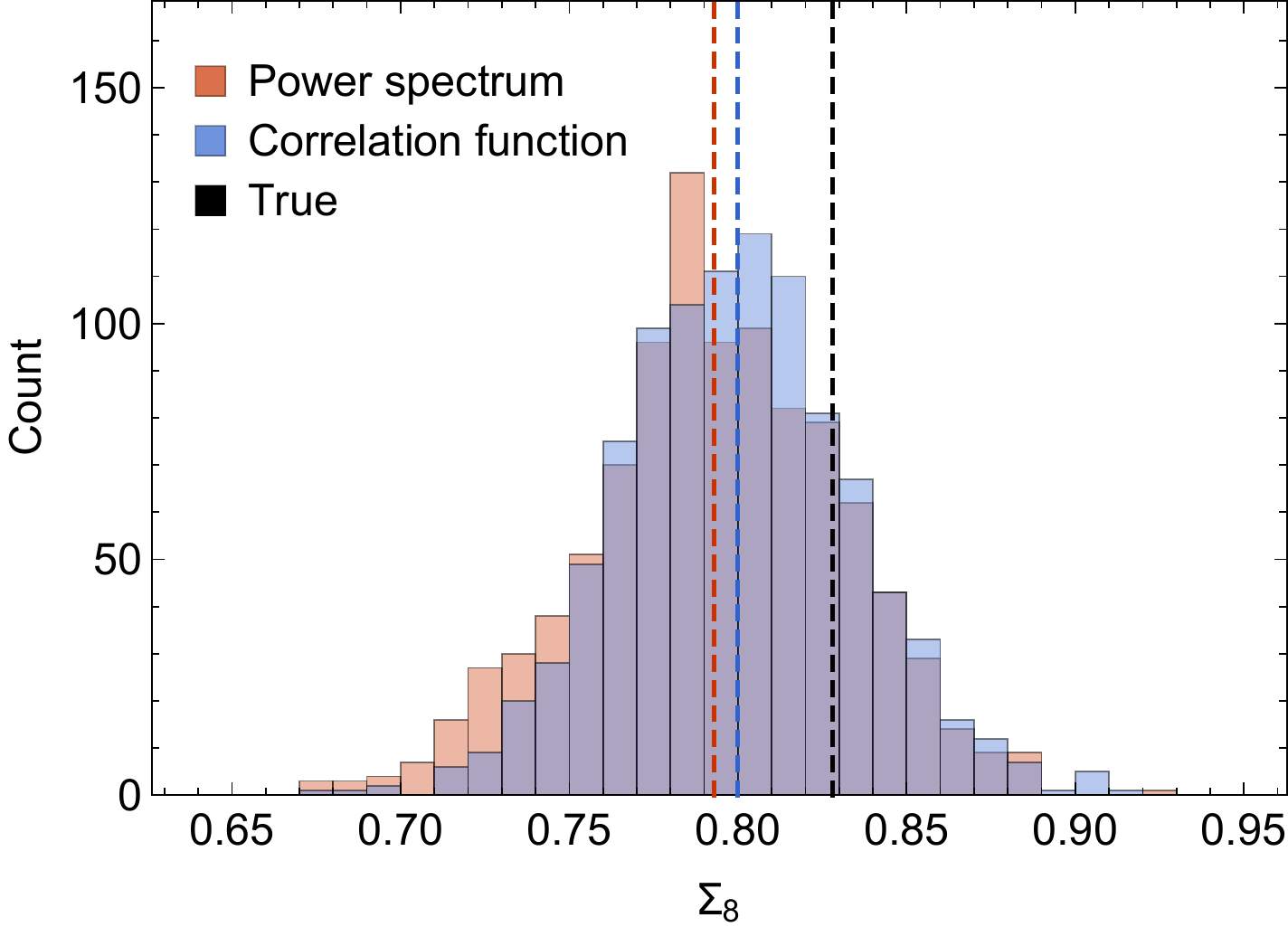}
\caption{We fit the 2PCF and the power spectrum to simulated, rather
  than real observed data, to check for systematic biases. The figure
  shows the distribution of $\Sigma_8\equiv\sigma_8(\Om/0.27)^{0.65}$
  inferred from 1000 realizations of mock shear catalogues in the
  fiducial cosmological model, using the two statistics. A
  cosmology-dependent covariance matrix is used. The average values
  are shown by the dashed vertical lines, and the true value is marked
  for comparison. Both statistics are biased low, by nearly the same
  amount.}
\label{fig:biases}
\end{figure}

Figure~\ref{fig:biases} shows the distribution of the 1000 best-fit
$\sigma_8(\Om/0.27)^\alpha$ ($\alpha=0.65$) values, using both the
2PCF and the power spectrum.  The true value in the fiducial model is
$\Sigma_8=0.828$, whereas the 2PCF and the power spectrum yield
average values of $\Sigma_8=0.800\pm0.001$ and
$\Sigma_8=0.793\pm0.001$, respectively.  The difference between these
values estimated from the two methods is $\delta\Sigma_8=0.007$,
which is nearly an order of magnitude smaller than required to explain
the difference ($\delta\Sigma_8=0.896-0.833=0.063$) found in \S~\ref{sec:verses} above.

On the other hand, we observe a systematic bias, with both statistics
under-predicting the true value by $\delta\Sigma_8=-0.03$. There are
two primary reasons for such a bias. First, the usage of
$\Sigma_8=\sigma_8(\Om/0.27)^\alpha$ as a parameter assumes a
power-law form for the degeneracy between $\Om$ and $\sigma_8$. But
the probability distributions deviate from a power-law, regardless of
the statistic \citep{hikage2019, lk2015}, causing the
maximum-likelihood estimators to be biased \citep{lk2015}.  Second, a
model with cosmology-dependent covariance is known to give a slightly
lower $\Sigma_8$ estimate compared to the true value
\citep{eifler2009}, because it gives more weights to low-$\Sigma_8$
cosmological models.

\subsection{Statistical fluke}
\label{sec:fluke}

\begin{figure}
\center
\includegraphics[width=7.5cm]{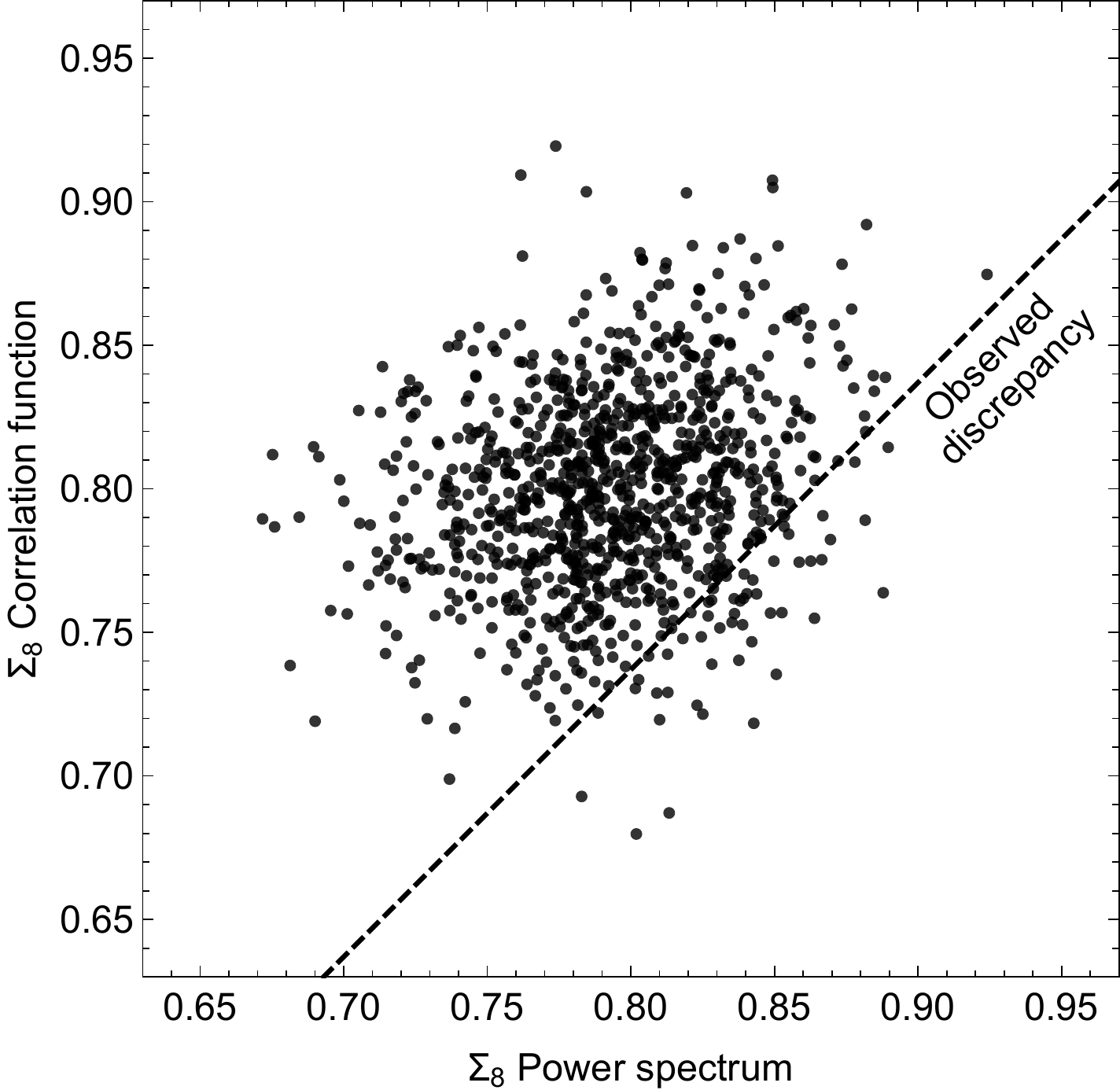}
\caption{The scatter plot of the $\Sigma_8$ values estimated from the
  2PCF and the power spectrum in each of the individual 1000
  realizations of the fiducial model. The discrepancy arising when
  fitting the CFHTLenS data (\S~\ref{sec:verses}) is shown by the
  dashed line. A discrepancy as large as the observed one is
  reproduced by chance in only 53/1000=5.3\% of the realizations.}
\label{fig:fluke}
\end{figure}

The possibility that the discrepancy is caused simply by a statistical
fluke should also be considered. We test the probability of this by
measuring the fraction of realizations of the mock shear catalogues
which yield a discrepancy equal to or larger than the observed
one. Figure~\ref{fig:fluke} shows a scatter plot of $\Sigma_8$
inferred from the two statistics in the 1000 individual realizations.
The correlations are visibly weak (with a correlation coefficient of
$R=0.22$). By counting the number of points with
$\Delta\Sigma_8\ge0.063$, we find that the chance of the observed
discrepancy among the realizations due to randomness is $5.3\%$.
Therefore, we conclude that a statistical fluke is not a particularly
compelling explanation: its likelihood is the same as the significance
of the original discrepancy (both at the $\approx2\sigma$ level).

\begin{figure}
\center
\includegraphics[width=7.5cm]{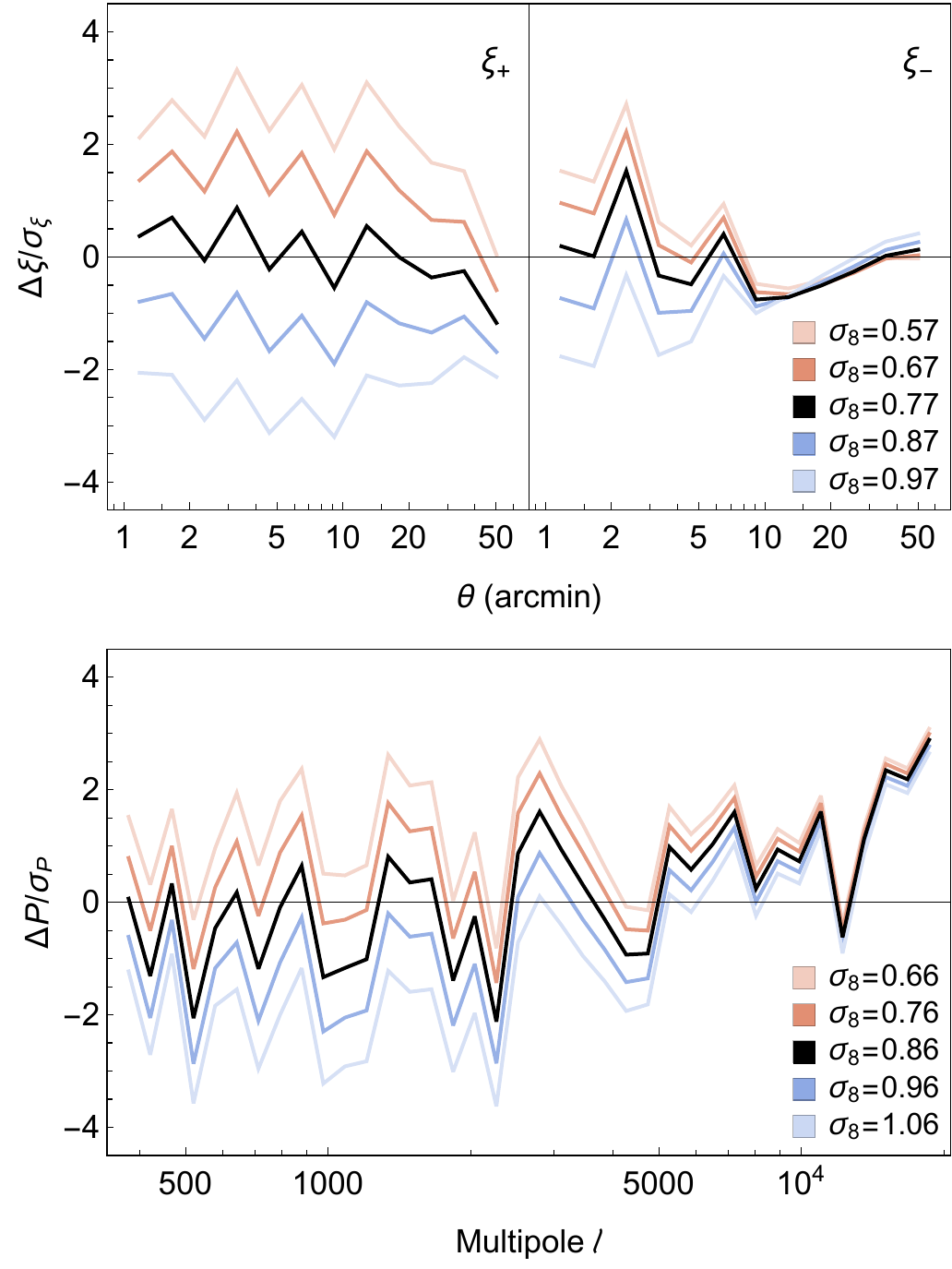}
\caption{The differences in the 2PCFs and the power spectra, in units
  of their standard deviations, between the CFHTLenS data and in
  ray-tracing simulations in a range of cosmologies. We use a fixed
  $\Om=0.3$ and a range of $\sigma_8$ values as shown in the
  legends. The best fits for $\sigma_8$ are shown in black. }
\label{fig:mock-compare}
\end{figure}

\clearpage
\newpage
\subsection{Shapes of 2PCFs and power spectra}

Finally, we check the profiles of the 2PCFs and the power spectra in
the real data, against those derived from $N$-body simulations. The
differences between the two are shown as a function of angular scale
in Figure~\ref{fig:mock-compare}. From the spreads and biases of the
residuals seen in this figure, we find that the 2PCF and the power
spectrum constrain $\sigma_8$ on different scales. For the 2PCF,
values from all scales of $\xi_+$ are similarly constraining, while
for $\xi_-$, smaller scales ($\lsim 5$ arcmin) are much more
important. For the power spectrum, the situation is the reverse: most
of the constraints come from relatively large scales ($\ell\lsim
4000$, corresponding to $\gsim 5$ arcmin).

A conspicuous feature in Figure~\ref{fig:mock-compare} is that the
real data has excess power on small scales ($5,000\lsim \ell\lsim
20,000$), compared to the simulations, by $\sim(1-3)\sigma$. Such an
excess power reveals the fact that the real data differs from what can
be generated by the $N$-body simulations.  More interestingly, this
excess small-scale difference only occurs in the power spectra,
without affecting 2PCFs, even though the two statistics are covering
the same range of angular scales ($1-60\,\mathrm{arcmin}$).

\begin{figure}
\center
\includegraphics[width=7.5cm]{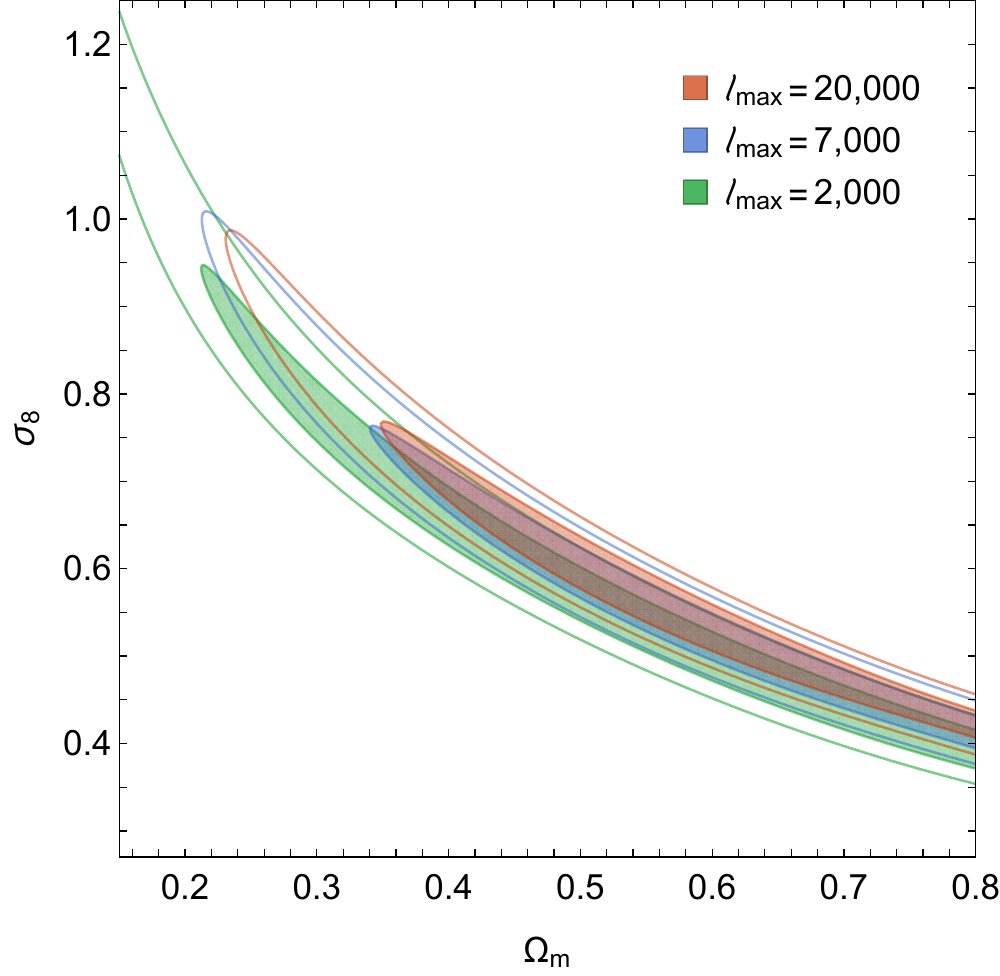}
\caption{68\% and 95\% joint confidence contours on $\sigma_8$ and
  $\Om$, using the power spectrum with different multipole
  cut-offs. $\ell_\mathrm{max}=20,000$ was used for our results in
  \S~\ref{sec:verses}, whereas the more conservative cut-offs at
  $\ell_\mathrm{max}=7,000$ and $\ell_\mathrm{max}=2,000$ were used by
  L15 and by \citet{hikage2019}, respectively.  Excising small scales
  reduces the best-fits $\sigma_8$ values; $\ell_\mathrm{max}=2,000$
  brings the results from the power spectrum in agreement with the
  2PCF.}
\label{fig:cutoffs}
\end{figure}

\begin{table}
\vspace{-2\baselineskip}
\centering
\caption{Constraints on $\Sigma_8$ from different statistics.}
\begin{tabular}{|c|c|c|c|}
\hline
Method & $\ell_\mathrm{max}$ & $\Sigma_8$ & $\alpha$ \\
\hline
\multirow{3}{*}{Power spectrum} & 20,000 & $0.896\pm0.038$ & \multirow{4}{*}{0.65} \\
  & 7,000  & $0.877\pm0.038$ & \\
  & 2,000  & $0.836\pm0.044$ & \\
\cline{1-3}
Correlation function & \diagbox[dir=NE]{}{} & $0.833\pm0.035$ & \\
\hline
\end{tabular}
\label{tab:cutoffs}
\end{table}

In order to investigate the impact of this excess power on the
parameter estimates, we repeated our analysis with a more conservative
choice of multipole cut-offs. In Figure~\ref{fig:cutoffs}, we show the
results from three multipole variations: (i)
$\ell_\mathrm{max}=20,000$ (our original choice), (ii)
$\ell_\mathrm{max}=7,000$ (used by L15 to avoid various artefacts on
small scales); and (iii) $\ell_\mathrm{max}=2,000$ (similar to the
value used by \citealt{hikage2019} to avoid the impact of the one-halo
term). The corresponding marginalised constraints on $\Sigma_8$ are
shown in Table~\ref{tab:cutoffs}. We find that a power spectrum
cut-off at lower $\ell_\mathrm{max}$ decreases the inferred $\Sigma_8$
values, and the power spectrum yields results consistent with the 2PCF
for $\ell_\mathrm{max}=2,000$. This demonstrates that the excess power
on small scales in the power spectrum explains the discrepancy with
2PCF.

\section{Discussion}
\label{sec:discussion}

\subsection{Field selection}
\label{sec:selection}

As mentioned in \S~\ref{sec:catalogue}, 25\% of the CFHTLenS data have
been found to fail a systematics test. In this section, we repeat our
previous analysis (\S~\ref{sec:results}), which was based on the 75\%
of the fields passing the systematics test, but this time we retain
the failed fields.

\begin{table}
\centering
\caption{Constraints on $\Sigma_8$ from different statistics, but
  including all fields (including those failing a systematics test).}
\begin{tabular}{|c|c|c|c|}
\hline
Method & $\ell_\mathrm{max}$ & $\Sigma_8$ & $\alpha$ \\
\hline
\multirow{3}{*}{Power spectrum} & 20,000 & $0.883\pm0.034$ & \multirow{4}{*}{0.65} \\
  & 7,000  & $0.860\pm0.034$ & \\
  & 2,000  & $0.818\pm0.040$ & \\
\cline{1-3}
Correlation function & \diagbox[dir=NE]{}{} & $0.853\pm0.036$ & \\
\hline
\end{tabular}
\label{tab:no-selection}
\end{table}

\begin{figure}
\center
\includegraphics[width=7.5cm]{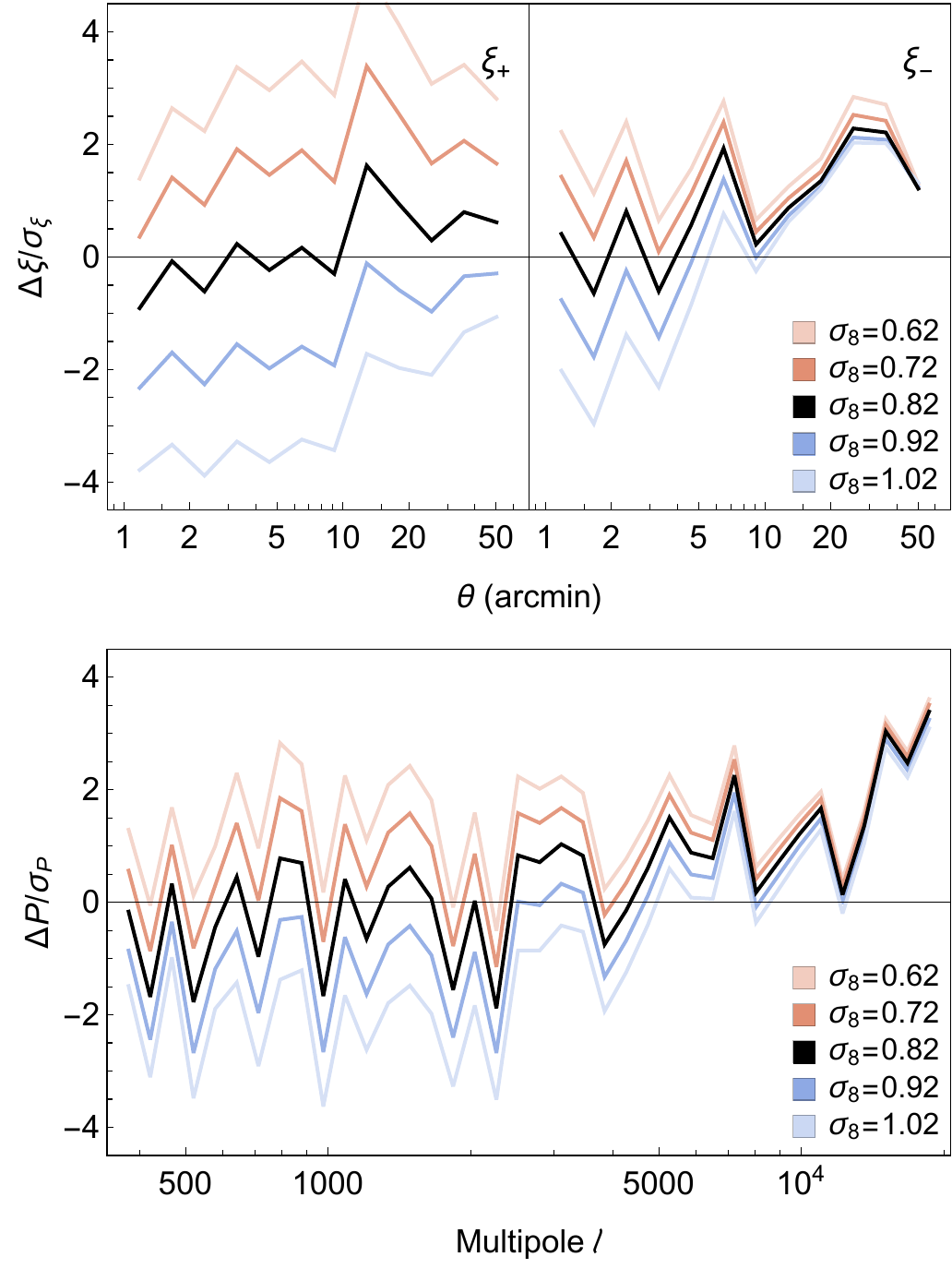}
\caption{The same as Figure~\ref{fig:mock-compare}, except the
  systematics test selection is not applied to the data (i.e. all
  fields are included). Note that the best fit value for $\sigma_8$
  (black) is different from Figure~\ref{fig:mock-compare}.}
\label{fig:no-selection}
\end{figure}

Table \ref{tab:no-selection} shows the marginalised constraints on
$\Sigma_8$ using all CFHTLenS data, and
Figure~\ref{fig:no-selection} shows the comparison between the
resulting 2PCFs and power spectra in the real data and those 
from the mock catalogues. We find that the excess power in the power
spectrum exists regardless of the systematics test selection.  We also
find that the inclusion of the data failing the systematics test
significantly alters 2PCFs, leading to excess correlations on large
scales at $1\sigma\sim 2\sigma$ level for $\xi_+$ and $\xi_-$.  This
is consistent with the reason these fields were excluded by the CFHT
team \citep{heymans2012}. Note that these systematic errors would
drive estimates of $\Sigma_8$ from the 2PCF higher.  Overall, we find
no evidence that residual systematics would help explain the observed 
difference between the 2PCF and the power spectrum.

\subsection{Cosmology-dependent vs. constant covariance}
\label{sec:covariance}

\begin{table}
  \vspace{-2\baselineskip}
  \centering
\caption{Constraints on $\Sigma_8$ from different statistics, but with
  a constant (cosmology-independent) covariance matrix.}
\begin{tabular}{|c|c|c|c|}
\hline
Method & $\ell_\mathrm{max}$ & $\Sigma_8$ & $\alpha$ \\
\hline
\multirow{3}{*}{Power spectrum} & 20,000 & $0.911\pm0.038$ & \multirow{4}{*}{0.65} \\
  & 7,000  & $0.890\pm0.040$ & \\
  & 2,000  & $0.847\pm0.047$ & \\
\cline{1-3}
Correlation function & \diagbox[dir=NE]{}{} & $0.847\pm0.042$ & \\
\hline
\end{tabular}
\label{tab:const-cov}
\end{table}

The results of using a constant covariance matrix are shown in
Table~\ref{tab:const-cov}. We find that the inferred $\Sigma_8$ values
are uniformly higher than those in the cosmology-dependent covariance
case, by $\Delta\Sigma_8\approx0.014$ in each case. The constant
covariance model underestimates the uncertainty of the statistics in
the cosmological models with low $\Om$ and/or $\sigma_8$ and
overestimates them for high $\Om$ and/or $\sigma_8$. Therefore,
in the case of constraining a marginalised parameter $\Sigma_8$, 
the constant covariance model has the tendency to overestimate the
likelihood of high $\Sigma_8$, yielding higher best-fit values.

Despite this overall shift to higher values, the differences between
the values of $\Sigma_8$ between the 2PCF and the power spectrum
remain similar. Therefore, we conclude that the assumed
cosmology-dependence of the covariance matrix does not explain the
apparent discrepancy.

\subsection{Reconstruction of the shape noise}
\label{sec:noise}

\begin{figure}
\center
\includegraphics[width=7.5cm]{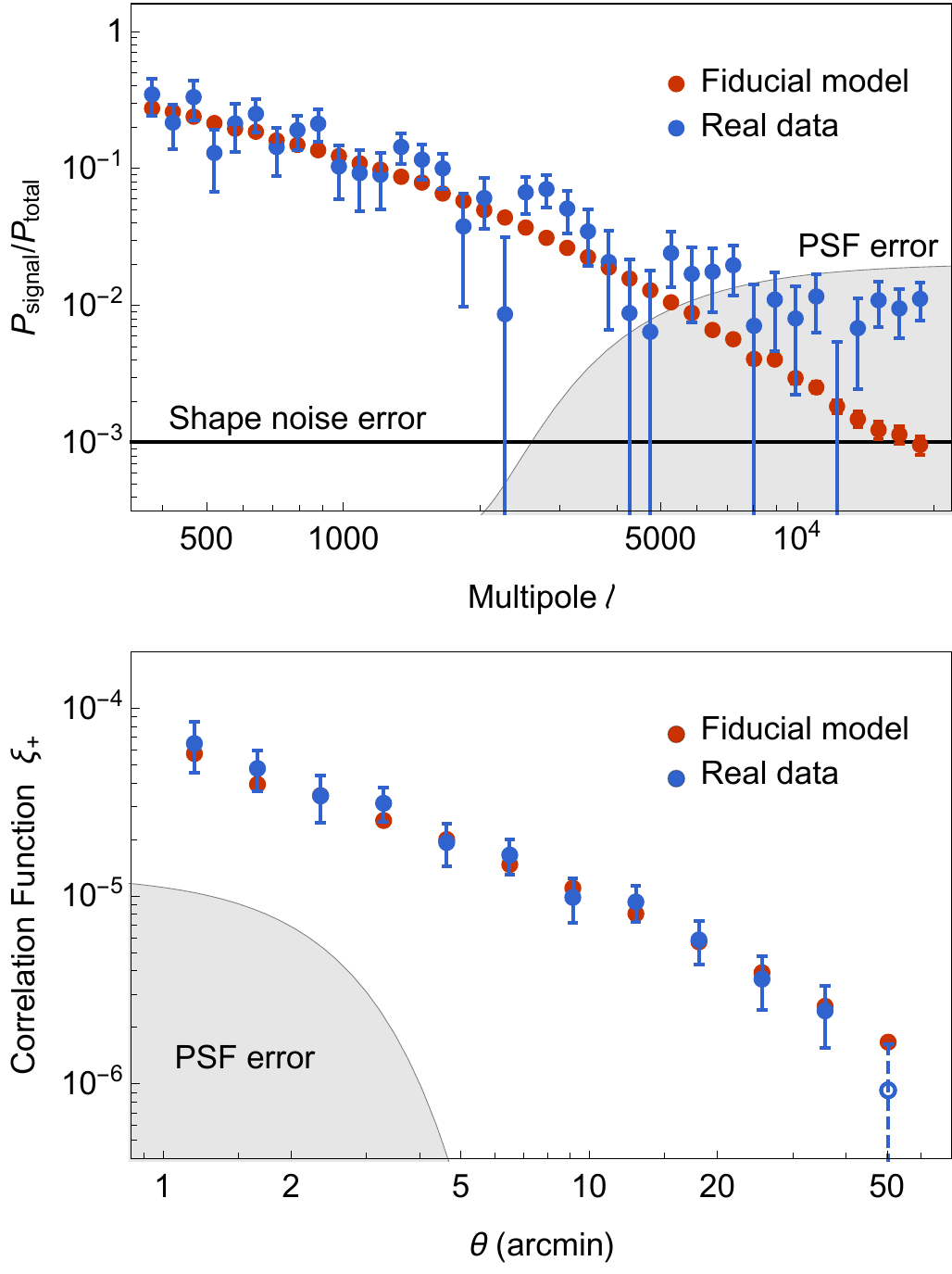}
\caption{The upper panel shows the contribution of the
    lensing signal to the total (=signal+noise) power spectrum in the
    fiducial model (red).  For comparison, we also show the difference
    between the power measured in the real data and the expected noise
    (blue; assuming the same shape noise power, and again divided by
    the same total power).
   The excess power shown in Figure~\ref{fig:mock-compare} also
   appears here at $\ell>5,000$. The estimated error in the shape
   noise is shown by the horizontal black line. The grey shaded region
   shows the possible range of residual power from PSF modelling
   errors (see \S~\ref{subsubsec:psf}).  The lower panel shows the
   residual correlation from PSF modelling errors compared with the
   2PCF $\xi_+$ (note that the shape noise is assumed to be uncorrelated and not to contribute).}
\label{fig:composition}
\end{figure}

The power spectra that we measured from the CFHTLenS Catalogue and the
mock catalogues consist of the lensing signal from the large-scale
structures, plus the random shape noise. Based on the fiducial
cosmological model, we find in Figure~\ref{fig:composition} that the
lensing signal power $P_\mathrm{signal}$ accounts for $\sim25\%$ of
the total power at $\ell=400$, but merely $\sim0.1\%$ at
$\ell=20,000$. Because the shape noise power $P_\mathrm{noise}$
dominates the total power, the accuracy of shape noise reconstruction
is very important on small scales. Here, we discuss two inaccuracies
in shape noise reconstruction as follows.

\subsubsection{Shape noise contamination}

We use randomly rotated ellipticities from the CFHTLenS Catalogue as
our proxy for the shape noise in the mock catalogues. An inaccuracy
from this method is that the true lensing shear in the data is
included as part of the shape noise. As a result, the shape noise
power in the mock catalogues is contaminated and overestimated. We
find that the dispersion of the lensing shear in the noiseless mock
catalogues is
\begin{equation}
\langle\left|e_\mathrm{signal}\right|^2\rangle=2\times10^{-4},
\end{equation}
while the dispersion of the observed galaxy ellipticities is
\begin{equation}
\langle\left|e_\mathrm{total}\right|^2\rangle=2\times10^{-1}. 
\end{equation}
Note that the random
rotations erase the correlations in the signal contribution to the
measured ellipticities, so that our adopted shape noise power is
overestimated by a constant, scale-independent factor of 
\begin{equation}
\delta P/P_\mathrm{total}=\langle\left|e_\mathrm{signal}\right|^2\rangle/\langle\left|e_\mathrm{total}\right|^2\rangle=10^{-3}, 
\end{equation}
as is shown in Figure~\ref{fig:composition}.  Comparing the shape
noise error to the power spectrum of the lensing signal in the real
data, we find it to be subdominant across all scales.

\subsubsection{Point spread function residual correlation}
\label{subsubsec:psf}

Point spread function (PSF) modelling also introduces inaccuracies to
the shape noise reconstruction. According to the CFHTLenS pipeline
\citep{erben2013}, PSFs at the locations of galaxies are estimated
based on nearby star images on the chip level, where a two-dimensional
second-order polynomial is used to model the local PSF
anisotropy. Then, the influence of PSF anisotropy is removed from the
galaxies. 
The PSF estimation method adopted in producing the CFHTLenS Catalogue 
was global third-order polynomials with some coefficients varying 
between CCDs \citep{miller2013}. The impact of PSF residual correlations
for CFHTLenS was studied by \citet{lu2017}.  While the 
above method was not specifically investigated, a similar method -- second-order 
polynomial fits on CCDs -- has been shown in \citet{lu2017} to produce residual 
correlations in lensing shear field of
$3\times 10^{-6} \lsim \delta\xi_+ \lsim 10^{-5} $ between $1\,\mathrm{arcmin}<\vartheta<3\,\mathrm{arcmin}$, which can be taken as an upper limit.

This residual correlation is subdominant to the shear-shear
correlation function (with a $\sim$10\% contribution on small scales; see the bottom panel
of Figure~\ref{fig:composition}).  Nevertheless, we find that it can have a non-negligible
impact on inferences from the power spectrum.
The Wiener-Khinchin theorem \citep{wiener1930} states that the 
correlation function and the power spectrum are related through a 
Fourier transform. Using this relation, we calculate the estimated 
error in the power spectrum by adding artificial noise to the mock 
catalogues according to the PSF residual correlation. In 
Figure~\ref{fig:composition}, we show the possible
range of PSF error assuming the residual correlation given by
\citet{lu2017} as an upper limit.
The PSF error is again subdominant to the total power, representing a
$\sim1\%$ contribution on small scales ($\ell \gsim 4,000$).  However,
unlike for the 2PCF, this exceeds the expected contribution of the
large-scale structure signal on these scales, and it is also
comparable to the excess power seen in the real data ($\delta
P/P_\mathrm{total}\sim 10^{-2}$). Therefore, we consider the PSF
residual correlation to be a possible explanation for the excess
power.  As discussed above (see Fig.~\ref {fig:mock-compare}), the
power-spectrum constraints are driven by somewhat larger scales,
but the excess small-scale power is still a plausible reason for
driving up the inferred $\Sigma_8$ value.

\subsection{Choice of spectral index in $N$-body simulations}
\label{sec:parameters}

We next explore the possibility that the choice of the spectral index
$n_\mathrm{s}$ may explain the excess power on small scales. In the
$N$-body simulations, we use a pre-defined set of cosmological
parameters (see \S~\ref{sec:mock}) based on WMAP
\citep{komatsu2011}. Our choice of $n_\mathrm{s}=0.96$ is lower than
the more recent and more accurate value $n_\mathrm{s}=0.965\pm0.004$
from Planck 2018 (TT,TE,EE+lowE+lensing) by up to $\sim1\%$.
With the fluctuation amplitude anchored at $8h^{-1}\,\mathrm{Mpc}$ by
$\sigma_8$, $n_s$ changes the logarithmic slope of the power spectrum
around this pivot scale. We now quantitatively study its impact on the
lensing signal power spectrum.

The CFHTLenS galaxy catalogue shows a median redshift of $\sim0.7$,
corresponding to an angular scale of $400\,\mathrm{kpc/arcmin}$. In
this case, the simulations correctly reconstruct the desired density
fluctuation at $30\,\mathrm{arcmin}$ or equivalently $\ell=800$. We
expect the $N$-body simulations to underestimate the power spectra on
smaller scales ($\ell>800$). In terms of $P_\mathrm{signal}$, for each
percent of increase in $n_\mathrm{s}$, the power would be increased by
$1.8\%$ at $\ell=5,000$ and $3.3\%$ at $\ell=20,000$. However, we find
that the lensing signal power in the real data is higher than in the
fiducial model by a factor of order unity or more at $\ell>5,000$ (see
Figure~\ref{fig:composition}). This is much greater than could be
attributed to our choosing an incorrect $n_\mathrm{s}$. We therefore
rule out this possibility.

\subsection{Intrinsic alignment}

We also consider intrinsic alignment as an explanation for the
excess small-scape power. Intrinsic alignment describes the phenomenon
in which ellipticities of galaxies are correlated not only through
gravitational lensing but also in real space, through the physics of
galaxy formation. It is a well-known contamination to weak lensing
measurements. \citet{ks2003} and \citet{heymans2004} have found that
modelling intrinsic alignment is crucial for  tomographic weak lensing
measurements using multiple redshift bins. Although a two-dimensional 
analysis, as in this work, is less affected. \citet{sb2010} and 
\citet{sifon2015} have found intrinsic alignment to alter the profile 
of 2D lensing power spectra significantly on small scales, for a 
fiducial medium-deep survey.

According to the theoretical estimates by \citet{sifon2015}, a
nonlinear alignment model would cause a $~50\%$ increase in
$P_\mathrm{signal}$ on the scale of $\ell=10,000$. Comparing this
value to the observed discrepancy of a factor of $\approx 3$ at
$\ell=10,000$ in Figure~\ref{fig:composition}, we find that intrinsic
alignment is able to explain only $\sim20\%$ of the excess power.

\subsection{Baryonic effects}

We finally briefly consider baryonic effects as a
possible reason for the excess power.  The impact of baryon physics
on the power spectrum and correlation function, under different
assumptions about star-formation and feedback, has been investigated
by several authors.  In particular, \citet{semboloni2011,semboloni2013} and
\citet{chisari2018} compared the power spectrum and the 2PCF from a
dark-matter-only reference simulation to these quantities in
simulations that include the cooling and condensation of baryons in
DM halos, as well as star formation and AGN feedback that drives gas
out of halos.  In general, they found that cooling enhances power on
the smallest scales, but on the scales of interest here,
corresponding to $\ell < 20,000$, or $k\lsim 3h\,\mathrm{Mpc}^{-1}$,
the baryon effects (especially when AGN feedback is included)
suppress both the power spectrum and the 2PCF $\xi_+$ by $\sim
1-10\%$. This difference is below the uncertainty of the power
spectrum measurements, and has the opposite sign.  We conclude that
baryonic effects are unlikely to explain the excess power on arcmin
scales.

\section{Conclusion}
\label{sec:conclusion}

Based on the same data from CFHTLenS, \citet{kilbinger2013} and
\citet{liu2015} derived constraints on $\Sigma_8$ that differ by
$10\%$ ($\sim2\sigma$).  We investigated this difference, motivated by
the fact that the lower value \citep{kilbinger2013} is in tension,
while the higher value \citep{liu2015} is in agreement with the CMB
corresponding constraints from the CMB.  We find that this discrepancy
originates from the difference in the statistics used in the two
studies: the 2PCF of the shear vs. the power spectrum of the
convergence. We have done a fair comparison of these two statistics
and reproduced a similar discrepancy ($\sim8\%$).  The discrepancy
exists regardless of whether we adopt a cosmology-dependent or
constant covariance matrices in our analysis.

By examining the scale-dependence of the 2PCF and the power spectrum,
we find that the power in the real data in the range
$5,000<\ell<20,000$ is significantly larger than in the $N$-body
simulations used to fit the data. We identify this excess small-scale
power as the main reason that the power spectrum yields an unusually
high $\Sigma_8$ (i.e. higher than in other WL studies). We find that
this excess power exists regardless of whether CFHTLenS fields that
fail a systematics test are included or not. When we chose a lower
cut-off $\ell_\mathrm{max}=2,000$ for the power spectrum, we recoved
agreement between the $\Sigma_8$ values inferred from the power
spectrum and the 2PCF.  This demonstrates that the small-scale excess
power is the reason for the discrepant $\Sigma_8$ values.

We speculate that two possible effects explain the excess power. 
First, we find that PSF residual correlations may introduced an
error in the power spectrum, causing an over-estimate by up to
$\sim10^{-2}P_\mathrm{total}$.  This is comparable to the excess power
seen in the real data at $\ell>4,000$, and thus a possible
explanation for this excess. Second, intrinsic alignment of galaxies
is expected to increase the amplitude of the lensing power spectrum by
$\sim$50\%, which could explain up to $\sim20\%$ of the excess power. 
In addition, we find that baryonic effect are expected to be too small 
and in the wrong direction to cause the excess power.

Overall, we conclude that the value of $\sigma_8$ found recently by
\citet{liu2015}, which is $\sim10\%$ higher than in other lensing
analyses and is in agreement with the corresponding CMB constraints,
is due to the excess small-scale power in the CFHTLenS data, which
affects the power spectrum (but not the 2PCF).  This excess is likely
a residual artefact, and therefore it does not resolve the tension
between the matter density fluctuations amplitudes measured in lensing
and in the CMB data. A more general lesson from our
results is that analysing the 2PCF and the power spectrum in tandem
could be a useful tool to discover (and to help control) systematic
errors.

\section*{Acknowledgements}

We thank Greg Bryan, Martin Kilbinger and Jerry Ostriker for useful discussions, and
acknowledge support from NASA ATP grant 80NSSC18K1093.

\bibliographystyle{mnras}
\bibliography{highsigma8}

\end{document}